\documentclass[runningheads]{llncs}    

\usepackage{graphicx}
\usepackage{xspace}
\usepackage[square,sort,numbers]{natbib}
\usepackage{float}
\usepackage{xcolor}
\usepackage{listings}
\usepackage{ragged2e}
\usepackage{booktabs}
\usepackage{tabularx}
\usepackage{colortbl}
\usepackage{rotating}
\usepackage{subfig}
\usepackage[linktocpage=true,hyperfootnotes=false]{hyperref}
\usepackage{cleveref}
\usepackage{enumitem} 
\usepackage{algorithm}
\usepackage{algorithmicx}
\usepackage{algpseudocode}
\usepackage{setspace}
\usepackage{eurosym}
\usepackage{xspace}
\begin{document}

\title{A Survey on Process Variants Meta-modelling Approaches}
%
%
\author{Lisana Berberi\inst{1}}
%
\authorrunning{L. Berberi}
%
\institute{Steinbuch Computing Center, Karlsruhe Institute of Technology, Karlsruhe, Germany \\
\email{lisana.berberi@kit.edu}\\
\url{https://www.scc.kit.edu}
}

\newcommand{\eg}{e.\,g.\@\xspace}
\newcommand{\ie}{i.\,e.\@\xspace}

\maketitle              
\begin{abstract}
This paper introduces the concept of process variants in process-aware information systems (PAIS) during the design-time phase, where multiple variants of a single process must be specified. Today's organizations have to manage multiple variants of a given process, such as multiple order processes or payment 
processes for a specific product or service they offer.
Traditional business process management tools lack in adequately capture and represent explicitly these variants.
Hence,  for more than a decade an array of approaches have been proposed to tackle this gap.
A reference or customizable process model has been introduced to model these variants collections in a way that each variant could be derived by
inserting/removing an activity according to a process context. This survey reviews current literature by providing an overview of meta-modelling approaches
that have been extended in order to capture the variations of business processes. Moreover, we give a comparative analysis of these approaches based on different criteria we identified from the inventory activity, providing insights into their strengths and limitations.
This paper concludes that current approaches to process variants meta-modelling provide a comprehensive view of the conceptual level of process variants and the control-flow process perspective. While some approaches go a step further by capturing variability in resources or specialization among activities/processes.

\end{abstract}

\section{Introduction}

In many process-aware information systems (PAIS) during design-time phase, many variants of the same process often have to be specified. Here, we introduce the basic notion of a process variant as follows.
A process model variant or shortly named process variant is an adjustment of a particular process to specific requirements building the process context \citep{Hallerbach2008}.
A process context is directly related to all the elements that comprise a business process.
This includes several contextual properties, such as e. g., process domain properties, control
flows, goals specified, resources assigned, organizational units associated, etc. Depending
on the process context type, different variants of our process are required, whereas the context is described by country-specific, order-specific, invoice-specific, and payment-specific
variables.
There exist two general approaches that allow the modelling of these generated variants: \emph{multi-model} and \emph{single model}.
The former one (e.g. variants in our motivating example) as classified by authors in \citep{Hallerbach2008}, which means that they are designed and kept separately resulting in data redundancy as often model variants are similar or identical for most parts. Furthermore, is far from trivial to combine existing variants to a new one (semi-)automatically. This solution is feasible only if few variants exists or if they differ significantly from each other. \par
Whereas, in the latter one, these variants might be expressed in a single process definition with the excessive use of XOR-Splits. The resulting processes are large, difficult to understand and to communicate and overloaded, and new process definitions still comprise of all the past processes definitions they should replace \citep{Lisana2018}. Moreover, it isn't possible to distinguish between normal and variant-specific branchings (\eg , our \emph{PayInvoice} process includes a decision to pay by bank transfer, \ie , perform activity \emph{Fill in the settlement info} if bank transfer choice is selected and if activity \emph{Request payment by bank transfer} is either performed or skipped, whereas in the model-side it's ambiguous and mixed with the "normal" process logic), unless these variant-specific conditions are marked and represented explicitly using special conventions \citep{Hallerbach2008}.\par
To address these shortcomings a significant research efforts have been triggered and thus an array of approaches have been published. Hence, is crucial to provide a list of these approaches and provide an comparison among them.
The remaining sections of this paper are as follows: section 2 gives a motivating example we use throughout of this paper, section 3 gives a state-of-the-art of process variants meta-modelling approaches, section 4 provides a comparative analysis of them and section 5 draws conclusions.

\section{Motivating example}
\label{sec:MotEx}

\begin{figure}[H]
\centerline{\includegraphics[width=\textwidth, height=4.8in]{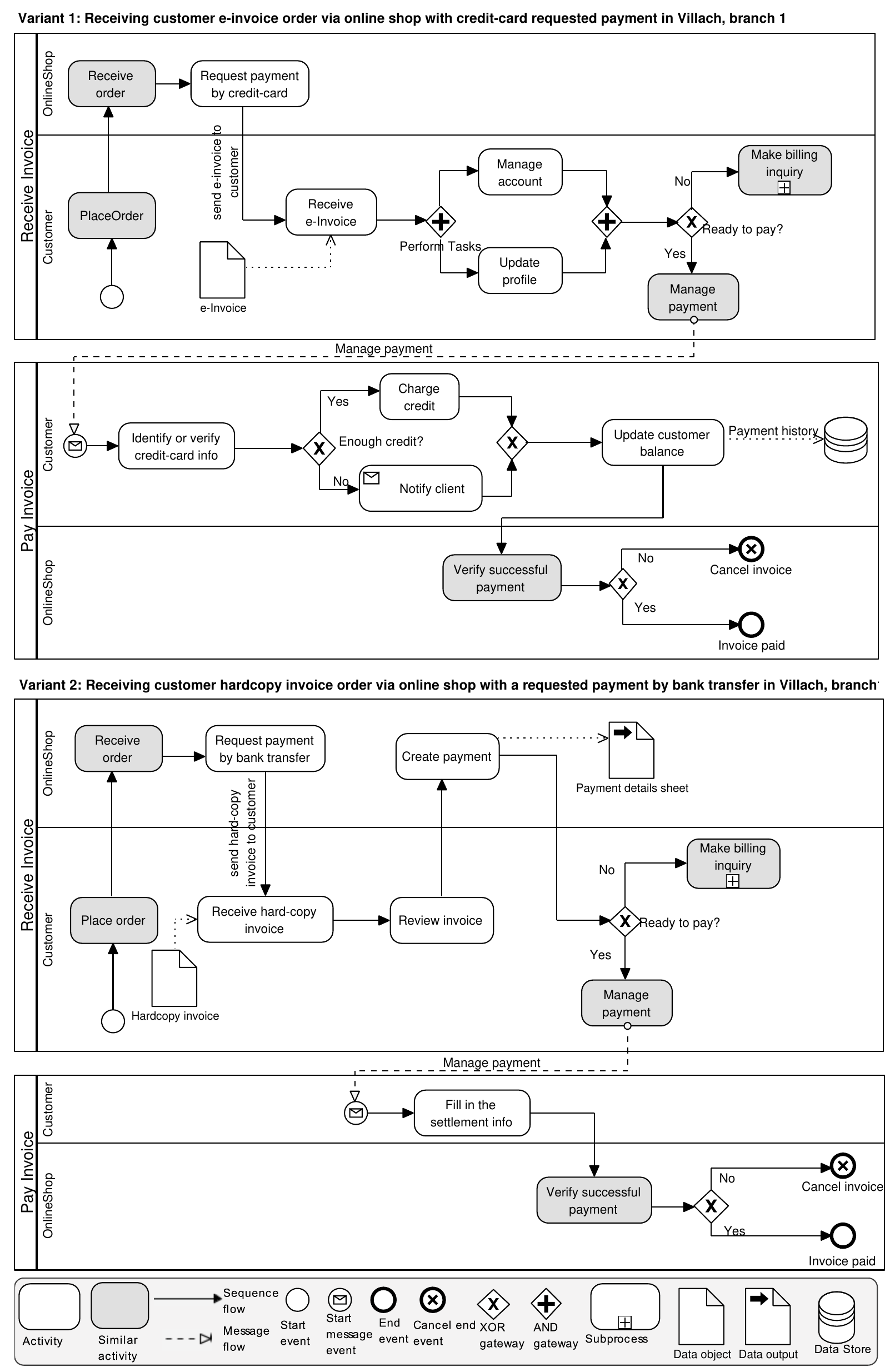}}	
	\caption{Process Variants (1)}
	\label{fig:Variant1}	
\end{figure}

\begin{figure}[H]
	\centerline{\includegraphics[width=\textwidth, height=6.5in ]{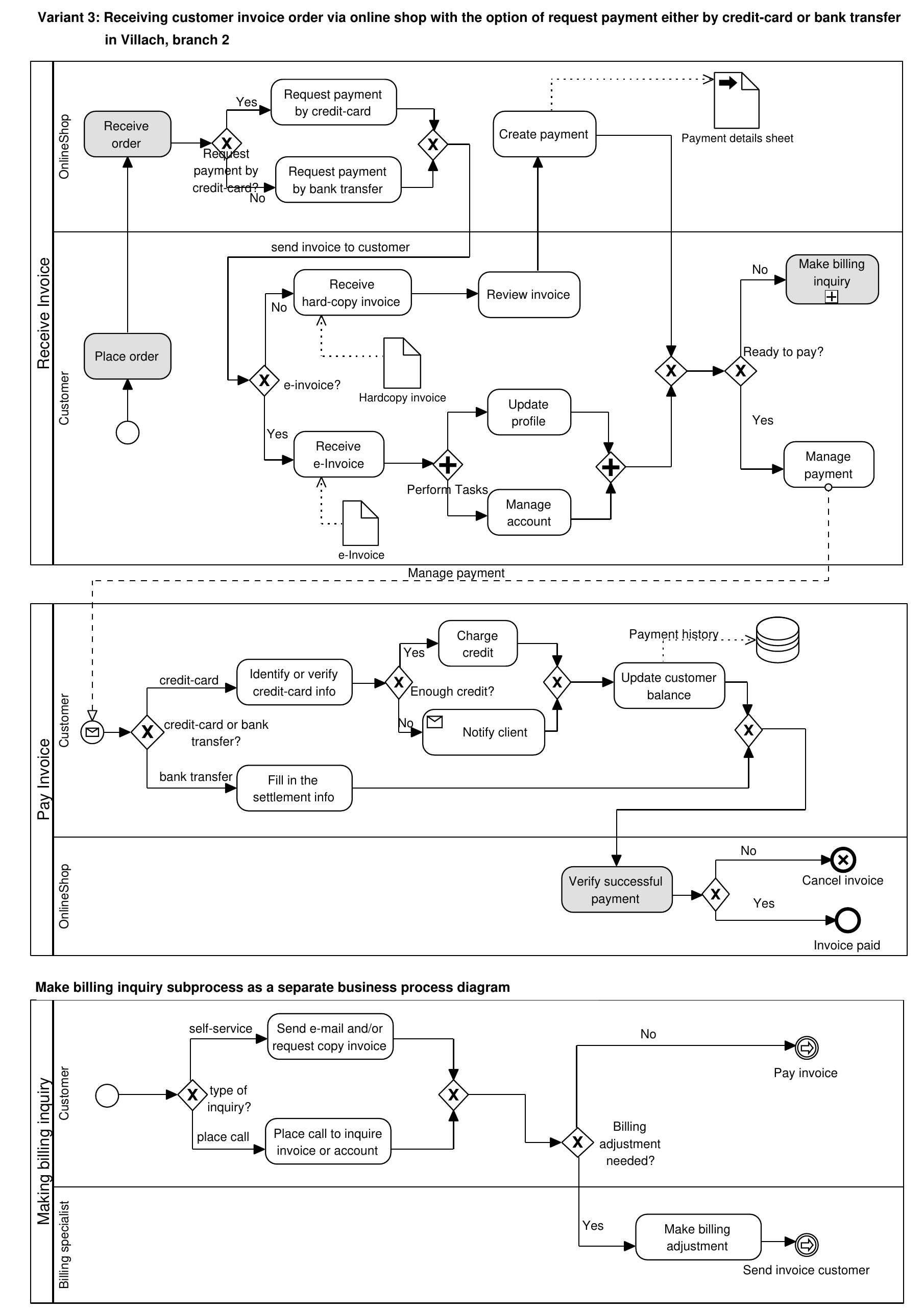}}		
	\caption{Process Variants (2)}
	\label{fig:Variant2}	
\end{figure}

\begin{figure}[H]
	\centerline{\includegraphics[width=1.25\textwidth, height=6.5in]{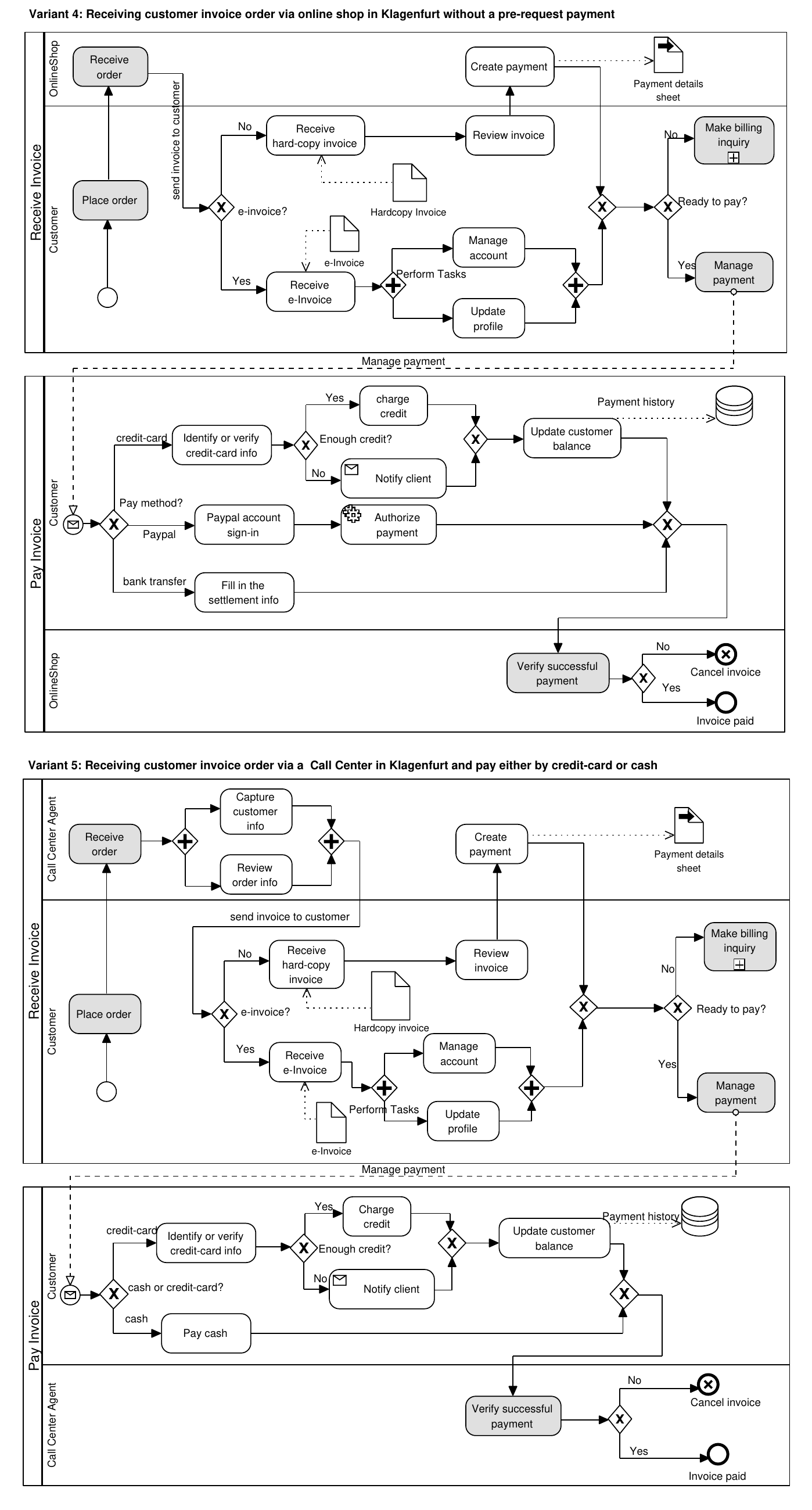}}		
	\caption{Process Variants (3)}
	\label{fig:Variant3}	
\end{figure}

Let's assume we have an illustrative example of a core business process, \eg , \emph{Processing Customer Invoice Payments} of a financial administration agency that is modeled as a collaboration between two processes named \emph{ReceiveInvoice} and \emph{PayInvoice}.
\emph{ReceiveInvoice} process consists of a set of activities that sends an invoice (either e-invoice or hard-copy) to a customer (buyer) with/without requesting a payment apriori for ordering its goods or services. Whereas, \emph{PayInvoice} process consists of a set of other activities that submit or complete with the payment (either by cash, bank transfer, credit-card or paypal) after a customer invoice is received.
We express these variants using BPMN notation \footnote{OMG: Business Process Model and Notation  http://www.bpmn.org/} which is now standardized by \citep{BPMNISO} to bridge the gap between business process design and process implementation.
We use this process with the set(family) of its variants as a running example throughout the paper.
Usually, dozens up to thousands of variants may exist of the same business process depending on different factors.
For example, in our running example variability is caused by the order customer choice (\eg , either via online shops or call centers) and/or method of invoice payment (\eg , either cash or credit-card) or invoice type (\eg , either hard-copy or e-invoice) of the designed process models in different branches of different cities.
Therefore, we represent these variants as shown in Figure~\ref{fig:Variant1}, Figure~\ref{fig:Variant2} and Figure~\ref{fig:Variant3}.
These five variants share some similarities highlighted with light gray, but they show differences, too.
A detail description about these variants is as follows.
All variants start with activity \emph{Place order} by a customer for ordering his/her goods and services. 

In variants 1-4 the order is received (\ie , activity \emph{Receive order} depicted with a rounded rectangle) via an online shop, whereas in variant 5 via a call center.
In the first three variants the processing of customer invoice payments is shown in different branches of the same city Villach, whereas variants four and five show how these processes are modelled in an agency located in a different city, \eg , in Klagenfurt.
After the order is received, either a request payment is followed (as in variant 1) or directly customer receives an e-invoice or hardcopy invoice. Then, \emph{Update profile} activity followed by a decision point (depicted with an X diamond) where subprocess \emph{Make billing inquiry} is executed if customer is not ready to pay otherwise activity \emph{Manage payment} (which is a common activity among all variants) is performed and the the flow is shifted to the second process \emph{PayInvoice} via a message start event (depicted with an envelope inside a circle). This subprocess deals with two types of inquiry by the customer: \emph{self-service} or via a \emph{place call} for further investigation related to invoice. If no billing adjustment are needed then the invoice can be paid executing the intermediate event to shift the control to process \emph{PayInvoice} otherwise activity \emph{Make billing adjustment} is performed by a billing specialist of the online shop to adjust billing items and afterwards the altered invoice is sent back to the customer. An expanded view of this subprocess is modeled as a separate business process diagram depicted in the bottom of Variant 3. \Cref{fig:Variant2}.
In process \emph{PayInvoice}, customer has different options to pay for his/her invoice, either by credit-card, (the credit is charged if customer has enough credit to its account), cash, bank transfer or PayPal, and finishes by \emph{Verify successful payment} activity. The order is successfully completed if the verification of the payment was successful otherwise is cancelled (e.g., due to insufficient credit amount).
Otherwise customer receives a notification about its insufficient credit amount following the cancellation of its invoice order (depicted with an x circle). 
After customer order is received activity \emph{Request payment by bank transfer} is performed. Afterwards, a hard copy invoice is sent to the customer (\emph{Receive hard-copy invoice}) generating a data object \emph{Hardcopy invoice} which serves as input object for activity \emph{Review invoice}. Then, a payment sheet named \emph{Payment details sheet} is generated as an output data object of activity \emph{Create payment} performed by an employee of the Online shop. In contrary to variant 1, the payment should be done by bank transfer as requested in the process \emph{ReceiveInvoice}. Accordingly, (\ie ,\emph{Fill in the settlement info}) is executed followed by activity \emph{Verify successful payment} otherwise the invoice is canceled if the bank transfer isn't settled with the right information. Based on the chosen invoice either payment by credit-card or payment by bank transfer is possible as expressed via the (\ie, decision point named \emph{Pay method?}). Again, after the verification of the payment the process completes with the event \emph{Invoice paid} for successful payments or event \emph{Cancel invoice} for unsuccessful ones. \\
In variant 4 a pre-request payment is not required by the online shop (instead a call center is used in variant 5), which means the decision is left to the customer after receiving an e-invoice or hard copy invoice. Here, possible payment methods are by credit-card, bank transfer or by third-party such as Paypal. 

\section{State-of-the-art of process variants meta-modelling}
\label{RelatedW_2}

This section provides an overview of cutting-edge meta-modelling approaches to capture variability of business process models for modelling and/or managing configurable processes. Instead of visualizing all proposed meta-models we show how our running example is represented in their solutions or through their case studies.
Therefore, we exclude from our analysis approaches that do not propose a meta-model for process variants. 
Furthermore, in the following subsections, from \crefrange{MMfPV1}{MMfPV4} we classify proposed approaches realized by means of different $\ll$\emph{variability mechanisms}$\gg$ suitable for business processes. 
A variability mechanism is defined as a technique for the derivation of process model variants from existing process models. We identified four different variability mechanisms: Inheritance and parameterization, Adaptation pattern-based, Template method pattern-based and Node configuration.

\subsection{Inheritance and parameterization variability mechanism}
\label{MMfPV1}
\begin{figure}[H]
\centering
\includegraphics[width=\textwidth, height=5.5in]{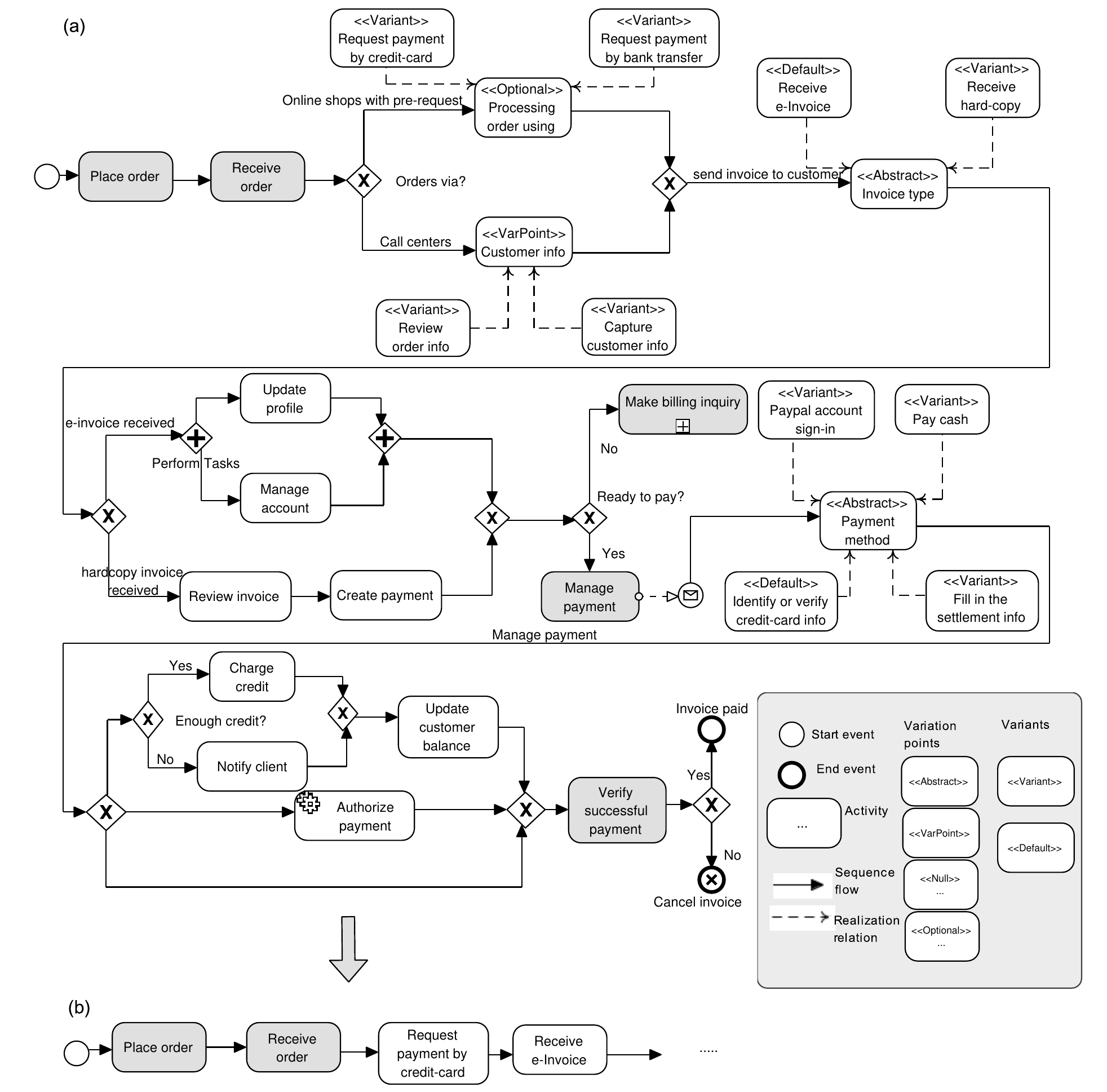}
\caption[Processing invoice payment example in PESOA-BPMN]{(a)Processing invoice payment example in PESOA-BPMN; (b) A customized model}
\label{fig:Approach1}
\end{figure}

As discussed by authors in \citep{puhlmann2005variability} these two variability mechanisms introduce: \textit{Inheritance} that allows for the replacement or addition of a model element, \eg , activity by the specialized one; \textit{Parameterization} that allows for controlling the behaviour of single execution step in a process by configuring the process with corresponding parameter values.
To introduce variability and configuration modelling to the processes in PESOA domains, \citep{Bayer05processfamily} proposed a conceptual model with variation points where fixed activities are marked with stereotypes applied in both UML ADs (Activity Diagrams) and BPMN. Their approach is called variant-rich process modelling (see Figure~\ref{fig:Approach1}).
The stereotype \emph{\(\ll\)VarPoint\(\gg\)} is assigned to activities of a process model in which variability can occur. An \emph{abstract} activity is represented by a variation point, such as "Customer info" that is specialized with one or more of the concrete variants (variants are inclusive). For example, "Review order info" and "Capture customer info" assigned with the stereotype \emph{\(\ll\)Variant\(\gg\)} are specializations of "Customer info". With the stereotype \emph{\(\ll\)Abstract\(\gg\)} are marked also abstract activities "Invoice type" and "Payment method" in our example where variation points are resolved by selecting only one of the concrete variants. \par
Figure~\ref{fig:Approach1} shows the process model for processing invoice payments in \emph{PESOA-BPMN} where activities have been marked as variation points with their variants, \eg , "Identify or verify credit-card info" marked as default activity of "Payment method", as being the most common choice in this process.
In this case, a variation point with the stereotype \emph{\(\ll\)Default\(\gg\)} represents the default variant.
Whereas, activity "Processing order using" is assigned with the stereotype \emph{\(\ll\)NULL\(\gg\)} to indicate its optional behaviour to one of the specialized activities annotated with \emph{\(\ll\)Optional\(\gg\)} stereotype. During customization variability in this point can be resolve by selecting one of its specialized variants such as "Request payment by credit-card" or "Request payment by bank transfer" or may completely be dropped from the process model. 
Accordingly, Figure~\ref{fig:Approach1} (b) shows an excerpt fragment of the configurable BPMN process model for the derived process variant of ordering e-invoice via online shops with a pre-request payment by credit-card.

\begin{figure}[H]
	\centerline{\includegraphics[width=\textwidth, height=5.0in]{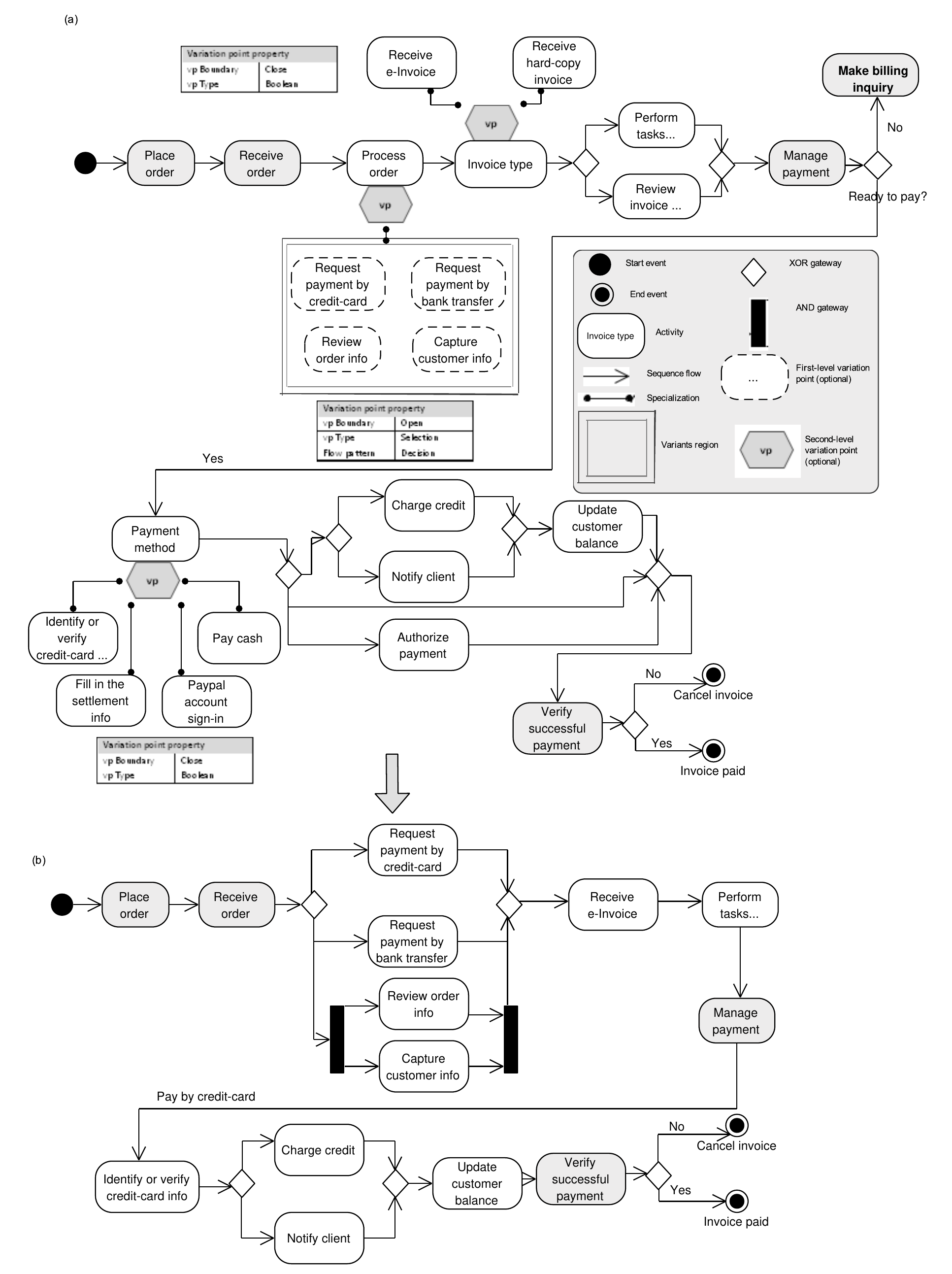}}		
	\caption[Processing invoice payment example in BPFM]{(a) Processing invoice payment example in BPFM; (b) A customized model}
	\label{fig:BPFM}	
\end{figure}

Another BPFM (Business Process Family Model) approach classified in this group is proposed by authors \citep{Moon2008} as a two-level approach.
They capture customizable process models using an extended version of UML ADs. They claim to have systematically conduct the realization of the variability in processes in different abstract levels in comparison with PESOA research project.
They represent variability using not only variation point and variant but also variation point type, boundary, and cardinality (see Figure~\ref{fig:BPFM}). At the first level, an activity can be defined as \textit{common} if it cannot be customized or \textit{optional} if it can be omitted during customization.

\begin{figure}[H]
	\centerline{\includegraphics[width=1.0\textwidth, keepaspectratio]{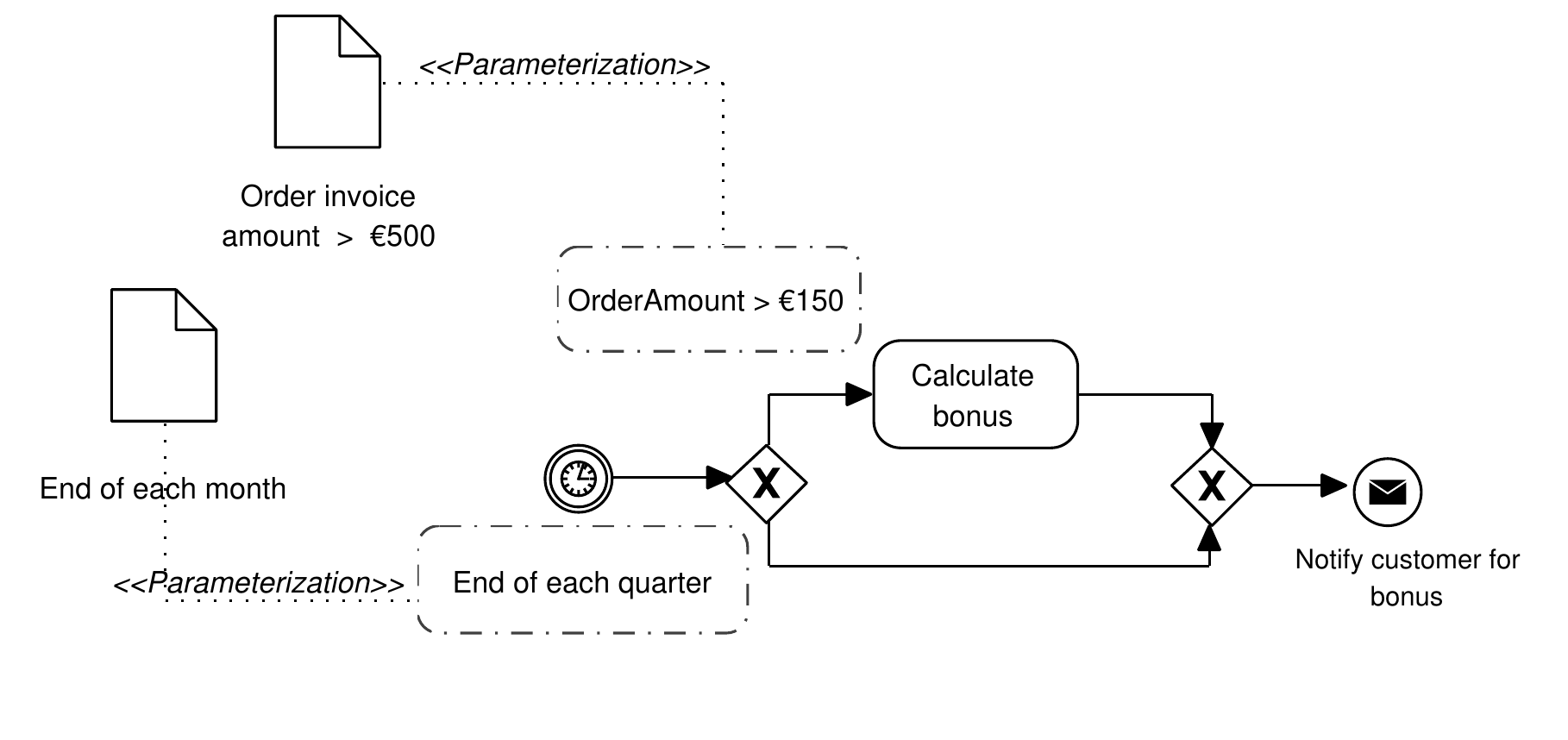}}		
	\caption[Parameterization of two different attributes in BPMN]{Parameterization of two different attributes in BPMN}
	\label{fig:PesoaPar}	
\end{figure}

Figure~\ref{fig:BPFM}(a) above, shows a customized model in which the first variation point has been customized to a decision between variants "Request payment by credit-card", "Request payment by bank transfer" and of another parallel execution of the other two remaining variants.
While the second and the last has been customized to one of the specialized activities, \ie , "Receive e-Invoice" and "Identify or verify credit-card info".\\
In Figure~\ref{fig:BPFM} (b) the first activity "Process order" indicate an open variation point of type flow with a decision pattern between activities "Request payment by credit-card", "Request payment by bank transfer", "Review order info", and "Capture customer info" in the associated variants region (depicted with a double rectangular).
The second level selects one of the specialized variants, \ie , concrete activities, which is represented by a variation point, \ie ,abstract activity. Variation points can be assigned only to activities.
Authors in \citep{Moon2008} identified three types of variation points (\textit{vpType}): \emph{Boolean} exactly one variant is selected from specialization; \emph{Selection} at least one variant is selected from a number of variants denoted with a cardinality (\eg , 1..2);  \emph{Flow}, a set of activities (expressed in a variant region) without a specified flow relation.
Whereas the second activity "Invoice type" indicate a boolean variation point, where only one of the variants can be selected.
And finally activity "Payment method" is of type boolean. Four variants are assigned to it, \ie , "Identify or verify credit-card info", "Pay cash", "Paypal account sign-in", and "Fill in the settlement info" and only one can be selected during customization. In our example, we don't indicate an activity of \emph{selection} type to express the fact that at least one or more variant may be assigned to it, as is the case of an OR-decision.\\
To abstract from the configurable process model and its variation points during configuration authors in \citep{Schnieders06} propose to use feature models (not presented due to page limits) in contrary to PESOA where authors in \citep{Bayer05processfamily} stated that the abstraction and transformation to derive process variants from a configurable process is out of their scope.
A feature model is represented graphically by one or more feature diagrams. 
A feature can be mandatory or optional (\ie , it can be deselected). It can be bound to other features via constraints (\ie , propositional logic expressions over the values of features).
\begin{figure}[H]
	\centerline{\includegraphics[width=1.0\textwidth, keepaspectratio, height=3.40in]{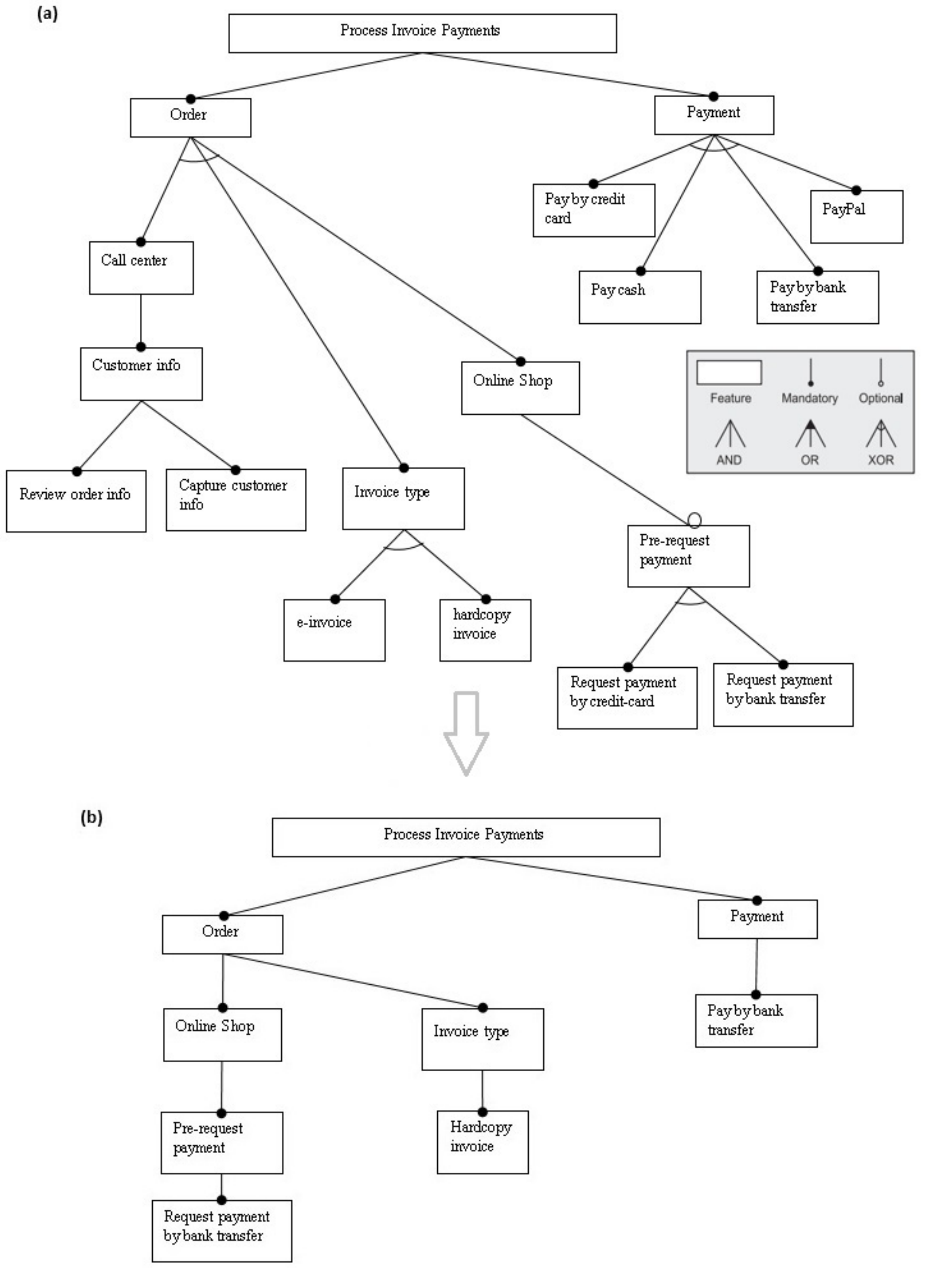}}		
	\caption[A feature diagram for invoice payments]{(a) A feature diagram for invoice payments (b) A possible feature configuration for invoice payments feature diagram}
	\label{fig:PesoaFeature}	
\end{figure}
For example, the subfeature "Request payment by bank transfer" of "Pre-request payment" must be deselected if the subfeature "e-invoice" of "Invoice type" of "Order" is not selected (see Figure~\ref{fig:PesoaFeature}). The relation between features and subfeatures is modeled using XOR (only one subfeature must be selected), AND (all the subfeatures must be selected), and OR (one or more might be selected). Even a feature is modeled as mandatory (arcs depicted with oval arrows) it can still be excluded if it has an XOR/OR relation with its sibling features. This is the case of the subfeatures of "e-invoice" which is mandatory and still is excluded in the configured feature diagram when subfeature "Pre-request by bank transfer" is selected.
However, a guidance is missing on how to perform the selection of a suitable set of features.\\
An association marked with the stereotype \emph{\(\ll\)Parameterization\(\gg\)} is used instead if a misinterpretation exist between the attribute and its corresponding element \citep{puhlmann2005variability}. The associations are used also to link data objects that contain the possible parameters to the grouping box that surrounds the attribute, see Figure~\ref{fig:PesoaPar}.     
This figure shows the parameterization of two different attribute where the lower one offers an alternative for the \emph{Datetime} attribute of the intermediate timer event. The alternative behaviour triggers the event at the end of each month whereas the default behaviour triggers the event at the end of each quarter.
In the upper side of this figure the \emph{ConditionExpression} attribute of a sequence flow is parameterized. The default parameterized attribute \emph{Amount} of an invoice order serves as a sentinel that activates the sequence flow if order amount is greater than a value (in this case, greater than \EUR{150}). Accordingly, the sequence flow is activated and a bonus is calculated for the customer. 
An alternative parameterization changes this attribute to activate the sequence flow if the order amount is greater then \EUR{500}. 

 Whereas, authors in \citep{Berberi2021_1000141036} proposed specialization and generalization notions between activities and so-called generic activities among process variants. We define generic activities (depicted with bold line rounded rectangular) as typical places where variation occurs among process variants. We have recently published how we semi-automatically derive the process variant hierarchy among activities and processes in this publication \citep{Lisana_10.1007/978-3-031-12670-3_8}. \par
\begin{figure}[H]
	\centerline{\includegraphics[width=\textwidth, keepaspectratio, height=5.9in]{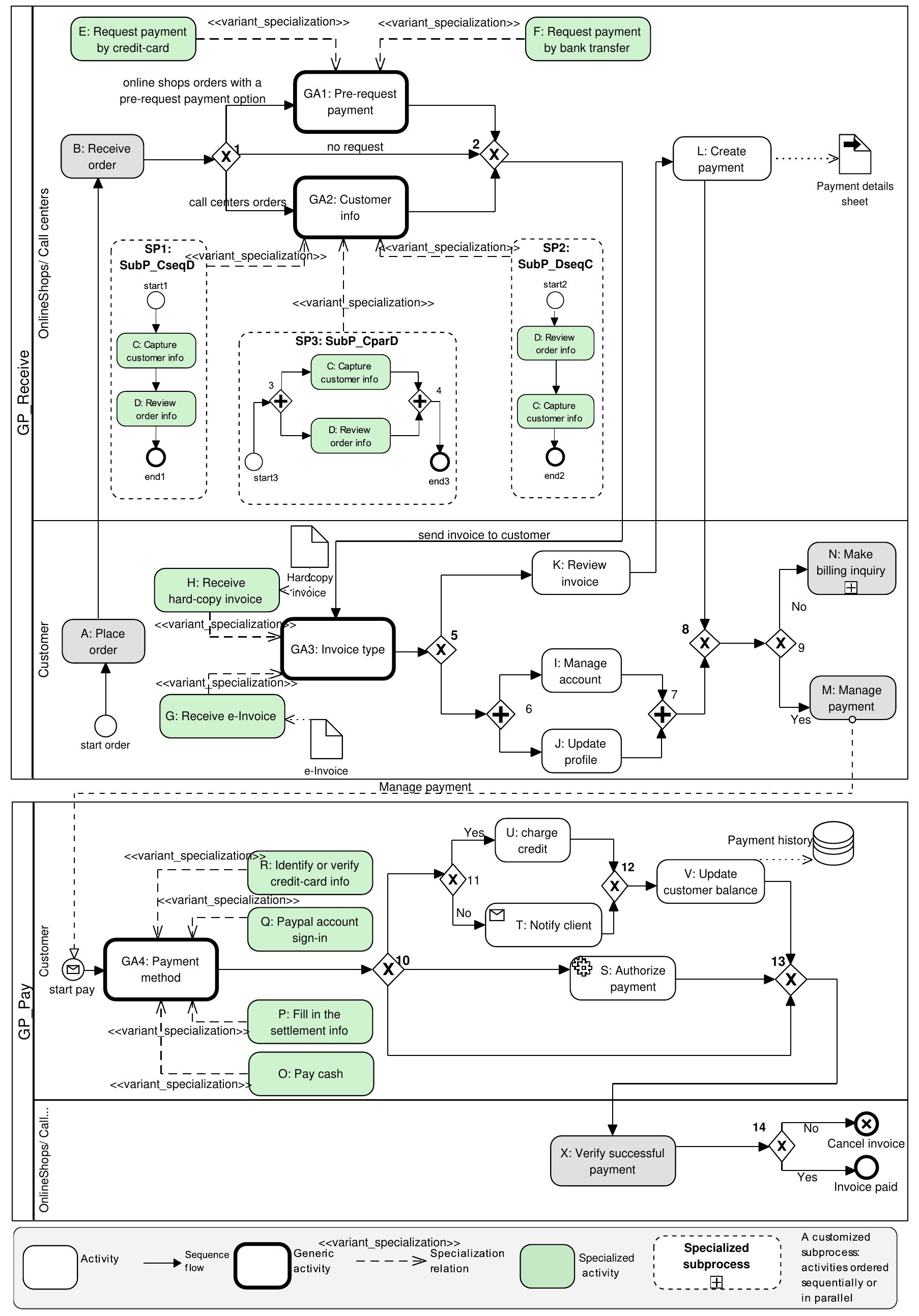}}		
	\caption[Invoice payment process model with generic activities]{Reference invoice payment process model with generic activities}
	\label{fig:GA}	
\end{figure}

Author annotates each of these connections between generic activities and specialized activities (either elementary or subprocess) with \emph{variant\_specialization} stereotype, to distinguish from the 'normal' sequence connector in process modelling (see Figure~\Cref{fig:GA}). For each realization of a generic activity to one of the specified activities every occurrence of a generic activity is substituted with the occurrence of respective activity from a concrete process. \par
In this figure we give an illustration of customizable invoice payment process as a reference (or so-called generic) process model with generic activities.
For example, activity "G: Receive e-invoice" is renamed by preceding letter 'G' as G: Receive
e-invoice, where G can be bound to one of the activities G1, G3, G4, G5 of the respective process variant (in variant two there is not such activity) it belongs. A generalization hierarchy can be generated from a meta-model of business process models which introduces the notion of generic activities which generalize a set of activities (\eg , pay by credit card, by check, or by third-party (PayPal) could all be generalized to an activity payment). Based on these given hierarchies of activities we can define generalization hierarchy of processes for the "process" dimension of a process warehouse.  This hierarchy can then be used to roll-up or drill down when analysing the logs of the executions of the various process variants and it makes it much easier to compare key-performance indicators between different variants at different levels of genericity.

\subsection{Adaptation variability mechanism}
\label{MMfPV2}
\begin{figure}[H]
	\centerline{\includegraphics[width=1.25\textwidth, height=4.1in]{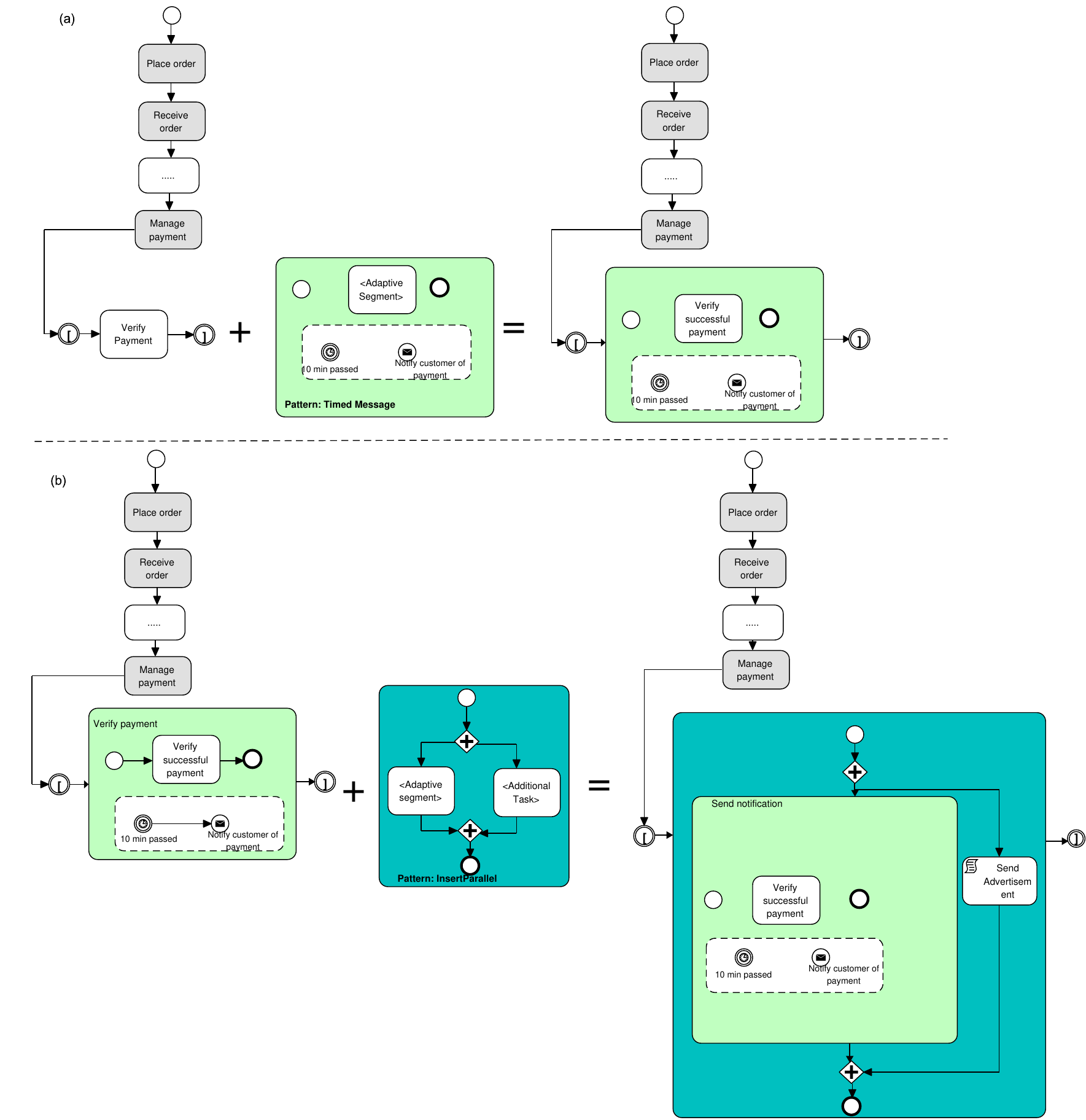}}		
	\caption[Pattern-based adaptation of a process in vBPMN]{(a) Time message-pattern adaptation of a process in vBPMN; (b) Insert parallel-pattern adaptation of a process in vBPMN}
	\label{fig:AdaptationPattern2}	
\end{figure}

Adapter design patterns are based on information hiding and inheritance like the Strategy design pattern\citep{puhlmann2005variability}. These patterns are used to represent processes using a combination of encapsulation and inheritance between process variants.
Instead, design patterns like 'Template Method' allows for controlling the behaviour of certain steps called 'placeholders' deferred to process runtime \citep{puhlmann2005variability}. Template approach is proposed for configuring a reference process based on a set of related business process models with an a-priori known variability \citep{Kulkarni2011} as well as on superimposed variants \citep{Czarnecki2005}.
Authors in \citep{Dohring2011} proposed the vBPMN(variant BPMN) approach to define the modelling of workflow variants by pattern- and rule-based adaptation in BPMN \citep{BPMN2.0Spec}.
Their approach consists of firstly, marking adaptive segments(variants) in a reference process, secondly, a BPMN2 adaptation pattern catalogue for realizing behaviour deviations and last, rules formulated in an event-condition-action (ECA) format applied to which adaptive segments and in what data-context.
\begin{figure}[H]
	\centerline{\includegraphics[width=\textwidth, height=4.2in]{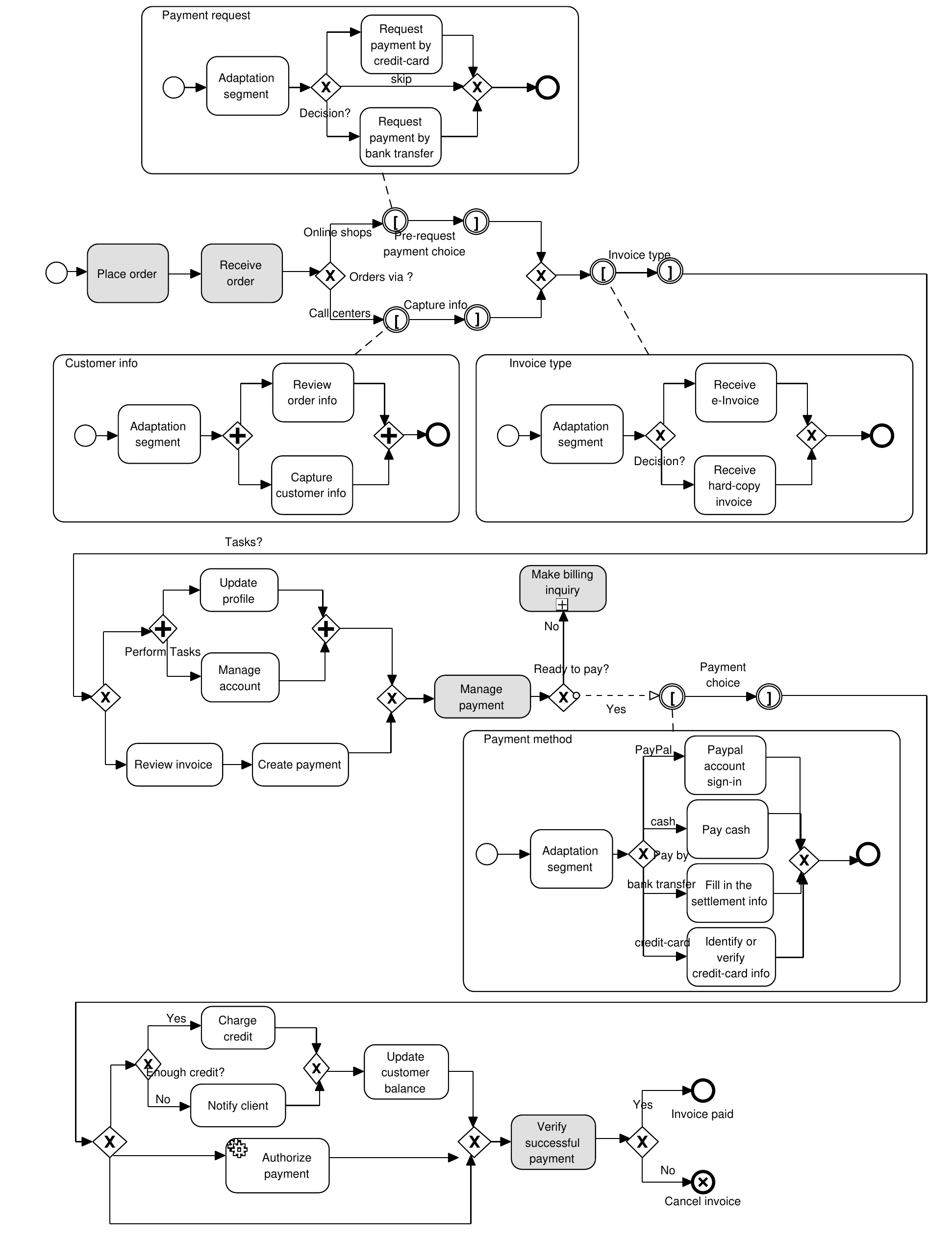}}		
	\caption[An adapted process model in vBPMN]{An adapted process model in vBPMN}
	\label{fig:AdaptationPattern}	
\end{figure}
To indicate the start and end of an \textit{"adaptive segment"} within a BPMN process definition two new nodes are introduced in vBPMN. 
An adaptive segment is structured as a single-entry, single exit point to facilitate the use of adaptation patterns.
Figure~\ref{fig:AdaptationPattern} shows our example of processing invoice payments with basically three adaptive segments, each of them marked between two opening/closing square brackets (depicted as intermediate events) indicating the modification of this segment using an adaptation pattern. 
Another annotation proposed by authors in \citep{Dohring2011} is by assigning a black diamond in the upper left corner of a single task. 
Each pattern consists of an implicit parameter \textit{$<$AdaptationSegment$>$} relating to which process segment it is applied on and the workflow engine gets notified whenever an adaptive segment is entered or left by explicitly annotating tasks. \\
To construct new variants the connection between the values of data-context variables and process tailoring operations needs to be established. This is achieved by formulating adaptation rules in an event-condition-action (ECA) format \citep{Dohring2011}. Each time a token enters an adaptive segment, the context variables are evaluated and the segment potentially becomes subject to immediate adaptations before continuing through the segment. \\
Another variant may be constructed if we annotate activity "Verify successful payment" as an "adaptive segment" to send an extra message notification of the verified payment to "special" customers (\ie , selected with a high status) for ordering their goods/services via call centers. This is achieved by adding a time-message pattern to the adaptive segment. 
Another adaptation can be annotated to this adaptive segmented to add an additional task in parallel with sending a message.
Parameterized patterns are applied to the adaptive segment by wrapping them around it as extensions as shown in Figure~\ref{fig:AdaptationPattern2}:\\
Rule \#{1}: realizes the send message event contextual facet for special customers placing their orders via call centres as shown in Figure~\ref{fig:AdaptationPattern2} (a).\\
Rule \#{2}: inserts an additional task for example "Send advertisement" for these type of customers as shown in Figure~\ref{fig:AdaptationPattern2} (b).


\textbf{RULE \#{1}}: ON verifyPayment{\_}entry IF orderVia="CallCenter" AND customerStatus="High"
THEN APPLY timedMessage(segment="Verification{\_}entry", HandlerTask=\\
						 "VerifyPayment", time=10 min)\\

\textbf{RULE \#{2}}: ON verifyPayment{\_}entry IF orderVia="CallCenter"
THEN APPLY  insert{\_}parallel(segment= "Verification{\_}entry", task="Send Advertisement")

Each adaptation rule has only one context factor, which uniquely assigns the rule to a distinct process variant.
However, there is no systematic way explained on how to mark these adaptive segments to capture variability on process models.\par
Another approach from authros in \citep{Hallerbach2008} is proposed, namely \textbf{Provop} (PROcess Variants by OPtions) for managing large collections of process variants in a single process. A set of change operations (\ie , insert, delete, modify and move) is used to describe the difference between \emph{basic process model} (\ie , the most frequently executed variant of a process family or process without a specific use case) and its respective variant model.
They identified requirements related to modelling of process variants, linking them to process context, executing them in WfMS, and continuously optimizing to deal with evolving needs.\\
Although different adaptation patterns such as: insert/delete, replace/move/swap process fragment or embedding the latter in loops, parallel or xor branches have been applied along to the entire process lifecycle \citep{Hallerbach2010, WEBER2011}, they are not yet sufficient to cope with complexity of process families \citep{Ayora2013}.
To this purpose, authors in \citep{AYORA2016} argued to address variability-specific needs of process families through change patterns that complements these adaptation patterns. Their approach namely \textbf{CP4PF} (Change Patterns for Process Families) comprise ten derived change pattern implemented in C-EPC for facilitating variability management in process families.
These CPs have been grounded empirically and validated in a real scenario through a case study 'check in process' in airline industry application with the aim to considerably reduce variability management effort.

\subsection{Template method pattern-based variability mechanism}
\label{MMfPV3}
\begin{figure}[H]
	\centerline{\includegraphics[width=1.5\textwidth, height=5.8in]{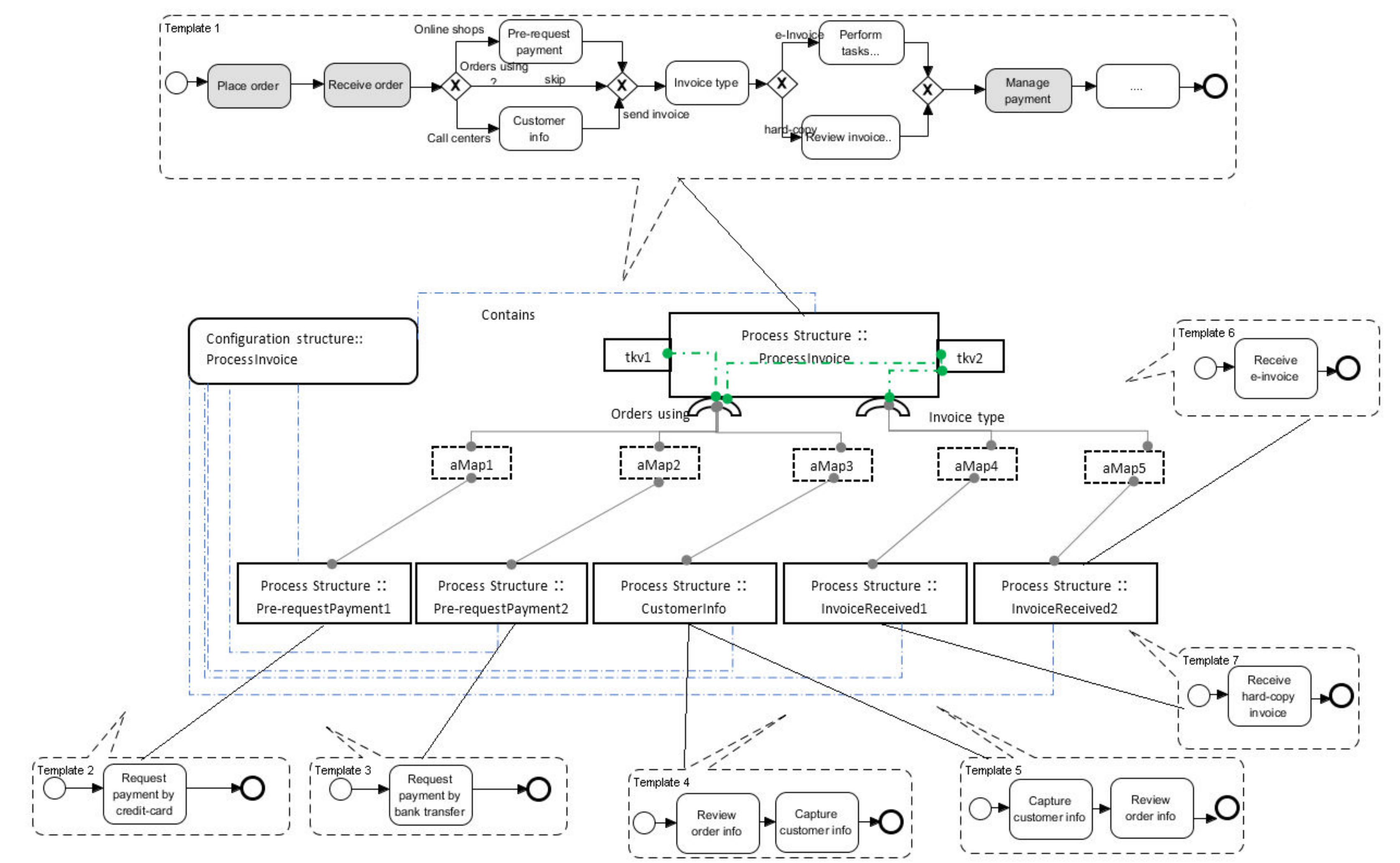}}		
	\caption[Process family of invoice payments]{Process family of processing invoice payments}
	\label{fig:TemplatePattern}	
\end{figure}

Design patterns like 'Template Method' allows for controlling the behaviour of certain steps called 'placeholders' deferred to process runtime \citep{puhlmann2005variability}. Template approach is proposed for configuring a reference process based on a set of related business process models with an a-priori known variability \citep{Kulkarni2011} as well as on superimposed variants \citep{Czarnecki2005}.
An essential BPMN meta-model is introduced by authors in \citep{Kulkarni2011} to capture the fixed behaviour (\ie , process structure) defined as a set of activities and events. A template constitutes of a control flow definition engaging fixed set of activities and events, \eg ,template1 in the following figure. 
A process structure is specified by multiple TKVs (\ie , a set of activity types). TKV is tuple $<$P, C$>$ where P is a set of abstract activities, \ie , placeholders, and C is set of concrete activities.
From TKV definition each placeholder is derived, which can be either an activity or an event. Variants assigned at a placeholder (\ie , part) are modeled explicitly using maps (\ie , a set of mappings describing fitment of a part at a placeholder) \citep{Kulkarni2011}
Figure ~\ref{fig:TemplatePattern} shows an excerpt of invoice payments process to illustrate variability and configurability. The process is defined as a control flow of set of activities A=\{Place order, Receive order, Pre-request payment, Invoice received, Perform tasks, Manage payment,..\} as depicted in Figure ~\ref{fig:TemplatePattern}. \\
A configurable process $P_{InvPayment}$ = \{$<$E, A, template1, D, tkv1 $>$\} of a process family PF =\{$<$ E, A, \{template1\}, D, TKV $>$\} where:\\
A = \{Place order, Receive order, Pre-request payment, Invoice type, Perform tasks, Manage payment,..\},\\
E = \{\},\\
template1= instance of essential BPMN meta model, and\\
TKV = \{tkv1, tkv2\} where tkv1 = \{P= \{Orders using\}, C= \{A $-$ \{Orders using\}\}\}, \\
tkv2 = \{P=\{Orders using, Invoice type\}, C=\{A $-$ \{Orders using, Invoice type\}\}\} and \\
D=data objects. 

Here, five \emph{activity maps} are defined, \eg , aMap1, aMap2, aMap3 are specializations of process structure "ProcessInvoice" if context \textit{Orders using} is selected. Whereas, aMap4 and aMap5 are specializations of process structure "ProcessInvoice" if context \textit{Invoice type} is selected. Some different behavioural variances through different configurations are as follows:-
\begin{enumerate}[label=(\alph*)]
\item configuration1=$<P_{InvPayment}$, \{Pre-requestPayment1\}, \{aMap1\}$>$ where \par
	  Pre-requestPayment1 =\{$<$\{\}, \{Request payment by credit-card \}, template2, D, $\phi$ $>$\}
\item configuration2 =$<P_{InvPayment}$, \{Pre-requestPayment2\}, \{aMap2\}$>$ where \par
      Pre-requestPayment2 =\{$<$\{\}, \{Request payment by bank transfer \}, template3, D, $\phi$ $>$\}
\item configuration3 =$<P_{InvPayment}$, \{CustomerInfo\}, \{aMap3\}$>$ where \par
      CustomerInfo =\{$<$\{\}, \{Review order info, Capture customer info \}, template4, D, $\phi$ $>$\}
\item configuration4 =$<P_{InvPayment}$, \{CustomerInfo\}, \{aMap3\}$>$ where \par
      CustomerInfo =\{$<$\{\}, \{Capture customer info, Review order info \}, template5, D, $\phi$\ $>$\}

\end{enumerate}
The same logic applies for the next context "Invoice type" as a specialization of process structure "ProcessInvoice".
A Configuration structure describes the entire configuration context in terms of parts that can be fitted at placeholders. It contains different process structures, in this case six as depicted in figure below.
Therefore, a configurable process with placeholders is a specialization of a template. The behaviours of configurable business process \emph{ProcessInvoice} can differ as different parts can be fitted at defined placeholder, \ie , abstract activity \emph{Pre-request payment} or \emph{Invoice type}.

\subsection{Node configuration variability mechanism}
\label{MMfPV4}

A node of a customizable process model, called configurable node is a variation point assigned to different customize options. 
Two main approaches fall in this group named Configurable Integrated Event-driven Process Chains (C-iEPC) and Configurable Workflows.
Authors in \citep{Gottschalk2008configurable,LAROSA2011,VanderAalst2005configurable,Rosemann:2007:CRM:1221586.1221839}
extended the EPC language for configuring a reference process model to capture multiple process variants in a consolidated manner.
Reference process models should be distinguished from so-called customizable process models.
A customizable process model is a concrete process model intended for a certain context, whereas a reference process model is intended to capture common behaviour or best practices of a family of process variants\citep{Fettke2003,Rosemann2003}.
In configurable workflows approac, authors in \citep{vanderAalst2006, Gottschalk2008configurable} presented C-YAWL(Configurable-YAWL), an extension of the executable process modelling language YAWL\footnote{www.yawlfoundation.org} where variation points in a process are configured using so-called \emph{ports}. Logic connectors (AND, XOR and OR) are integrated in each task in the form of a split (for the outgoing arcs) and a join (for the incoming arcs). In C-YAWL, like in C-EPC each feasible port variation is presented with a process fact. A C-EPC is an EPC in which functions and connectors can be marked as "configurable". A modeler can derive an individualized EPC from a C-EPC by selecting a possible variant for each configurable element\citep{LAROSA2011}.
As this approach doesn't present a meta-modelling solution for variants it's outside of the scope of this literature review section for a further discussion.
In C-iEPC approach configurable nodes might be activities, gateways, events as well as objects and resources presented with a meta-model.

For each configurable node one customization option is selected to achieve customization.
Configurable roles and configurable objects have two dimensions: optionality and specialization. 
If a configurable role (object) is "optional" (OPT), it can be restricted to "mandatory" (MND), or switched "OFF" to be removed from the process. If it is "mandatory" it can only be switched "OFF" \citep{LAROSA2011}.
There are some options for every configurable node, such as \textit{off} option which means node(s) does not appear in the customized model or \textit{on}, node(s) is being kept in the customized model.
Therefore, configurable nodes indicate the differences between process variants.
These variations in the extended notation, namely C-iEPC, are captured in the way roles and objects are assigned to activities. To maintain control-flow, resource and object perspectives synchronized is essential to prove the correctness of the individualized process model. \par

\begin{figure}[H]
	\centerline{\includegraphics[width=\textwidth, height=4.5in]{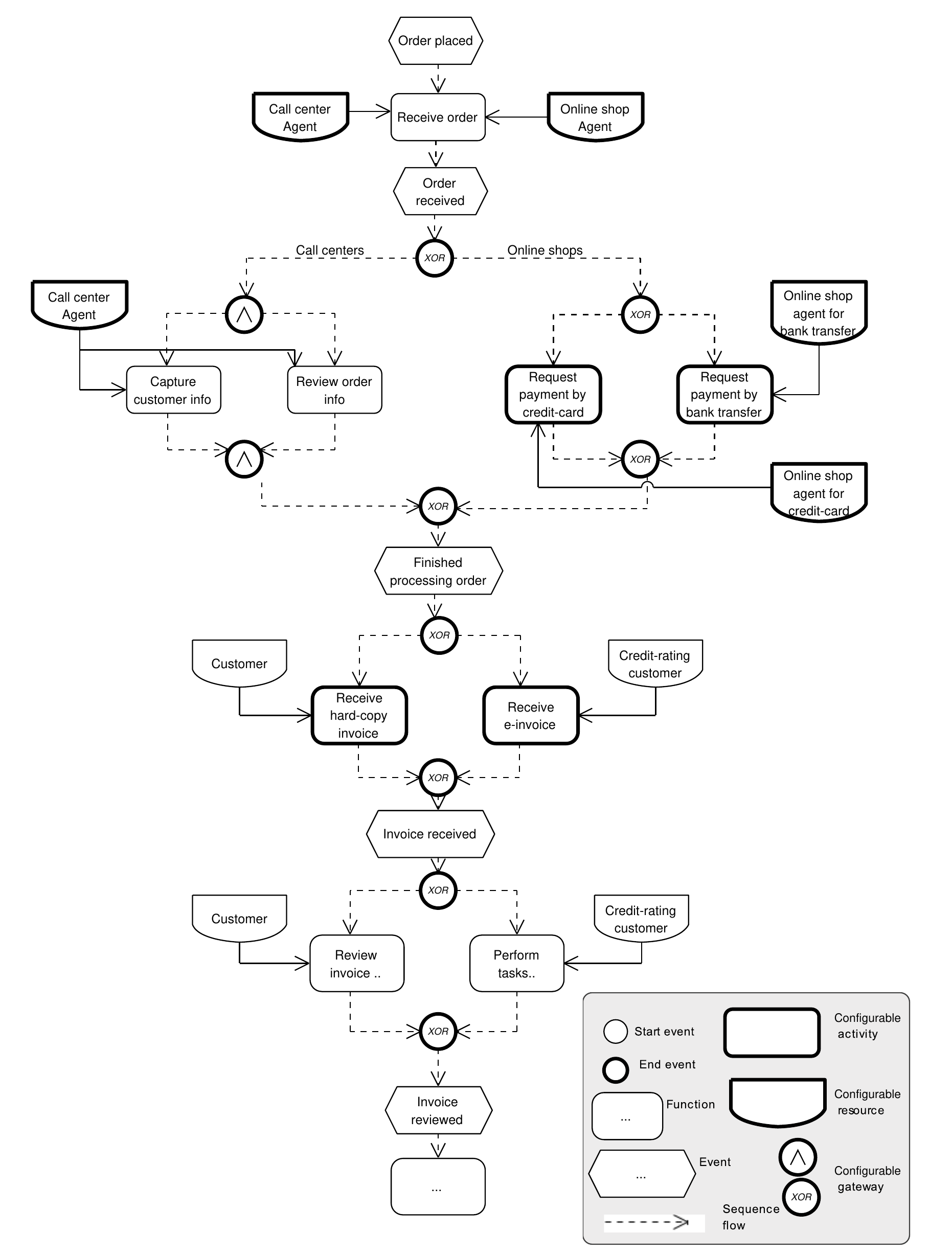}}		
	\caption{The C-iEPC model representing all invoice payment variants}
	\label{fig:NodeConfiguration}	
\end{figure}

The number of an OR outgoing flows (if it is a split gateway) or the number of its incoming flows (if a join) can be restricted to any combination (\eg , two flows out of three), including being restricted to a single flow, in which case the gateway disappears \citep{LAROSA2011}. \\

\begin{figure}[H]
	\centerline{\includegraphics[width=\textwidth, keepaspectratio]{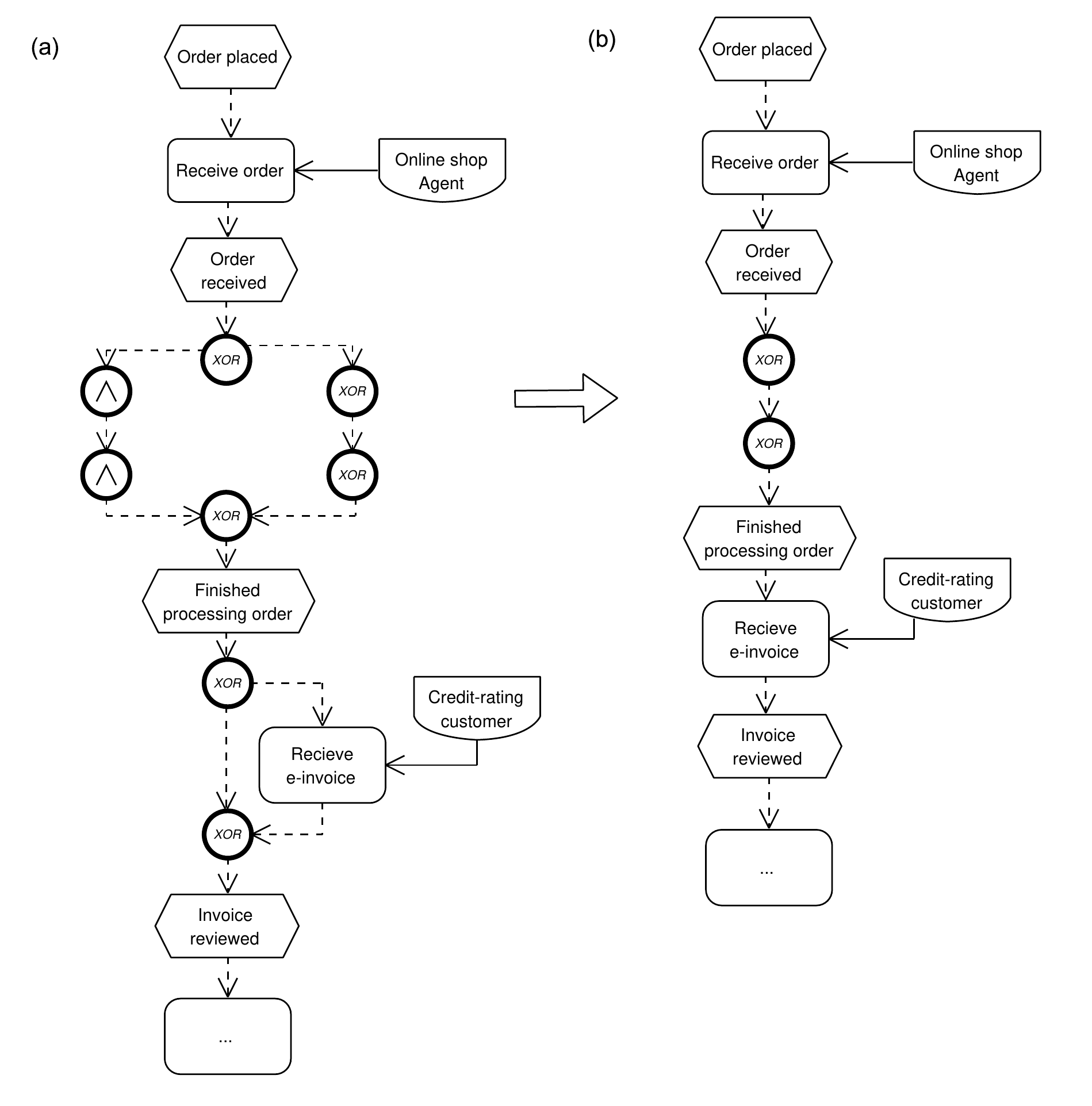}}		
	\caption[An individualized process model]{The application of the individualization algorithm to a fragment of processing customer e-invoice process model of Figure~\ref{fig:NodeConfiguration}}
	\label{fig:NodeConfiguration2}	
\end{figure}

Our case study of processing customer invoice payments is shown in Figure~\ref{fig:NodeConfiguration} which captures all five variants modeled as separate process models (see Section~\ref{sec:MotEx}) into one single process model.
Here, activities and gateways (\ie , variation points) are marked as configurable with a thicker border. 
Configurable gateways can be customized to an equal or more restrictive gateway. A configurable OR can be restricted to an XOR or to an AND gateway or can be left as a regular OR (no restriction). 
For example, we can capture the choice of processing orders via online shops or call centers by customizing the first XOR-split in Figure~\ref{fig:NodeConfiguration} or we can postpone the decision till runtime.
If the choice is "online shops" we can restrict this gateway to the outgoing flow leading to the event "Invoice reviewed". As a result, the branch starting with the sequence flow "Call centers" is removed, and vice versa. 
Configurable activities can be kept on or switched off. If switched off, the activity is simply hidden in the customized model. In addition, they can be customized to optional. The choice of whether to keep the activity or not is deferred until runtime. For example, the function "Request payment by credit-card" and "Request payment by bank transfer" are configurable nodes in Figure ~\ref{fig:NodeConfiguration}; thus, we can switch them off for those orders received via online shops for which a pre-request payment is not required.
Configurable elements might be resources (called roles in C-iEPCs) and objects, too. 
Authors in \citep{LAROSA2011} propose to use logical gateways so-called range connector (\ie , XOR, OR and AND) that allow any combination of the resources and objects connected to activities modeled by a pair of natural numbers, \eg , lower bound (2) and upper bound (5), which means at least 2 and at most 5 resources. 
For simplicity, Figure~\ref{fig:NodeConfiguration} depicts only three resources marked as configurable nodes with a thicker border meaning that during customization they can be configured to one of the specialized resources. However, it's out of our scope to demonstrate how the specialization of resources assigned to activities is achieved.\\
A formalized algorithm with a proven theory is presented to guarantee the correctness of the individualized process model (\ie ,an iEPC with non relevant options being removed) derived from a configurable process model with respect to a valid configuration.
Accordingly, functions "Capture customer info", "Review order info", "Request payment by credit-card", "Request payment by cash", "Review invoice" have been switched \emph{OFF} and thus they have been replaced by an arc. The resulted model is shown in Figure~\ref{fig:NodeConfiguration2} (a) whereas (b) is the result after applying the last step of the individualized algorithm where SESE control-flow connectors are replaced with arcs (this may result in consecutive events and consecutive functions to be interleaved with events have to be removed).
C-iEPCs do not provide any execution support, they are formally defined in \citep{LAROSA2011}.
An iEPC model is derived from a C-iEPC using an individualization algorithm. In the customized model all nodes that are no longer connected to the initial and final events via a path are removed and the remaining nodes are reconnected to preserve (structural and behaviour) model correctness. 
To capture domain properties and their values a questionnaire is linked to configurable nodes of C-iEPC, supported by Synergia\footnote{www.processconfiguration.com} and Apromore\footnote{www.apromore.org} toolsets. The resulting customized models have been validated via a case study in the film industry.\\

\begin{figure}[H]
	\centerline{\includegraphics[width=\textwidth, height=4.0in]{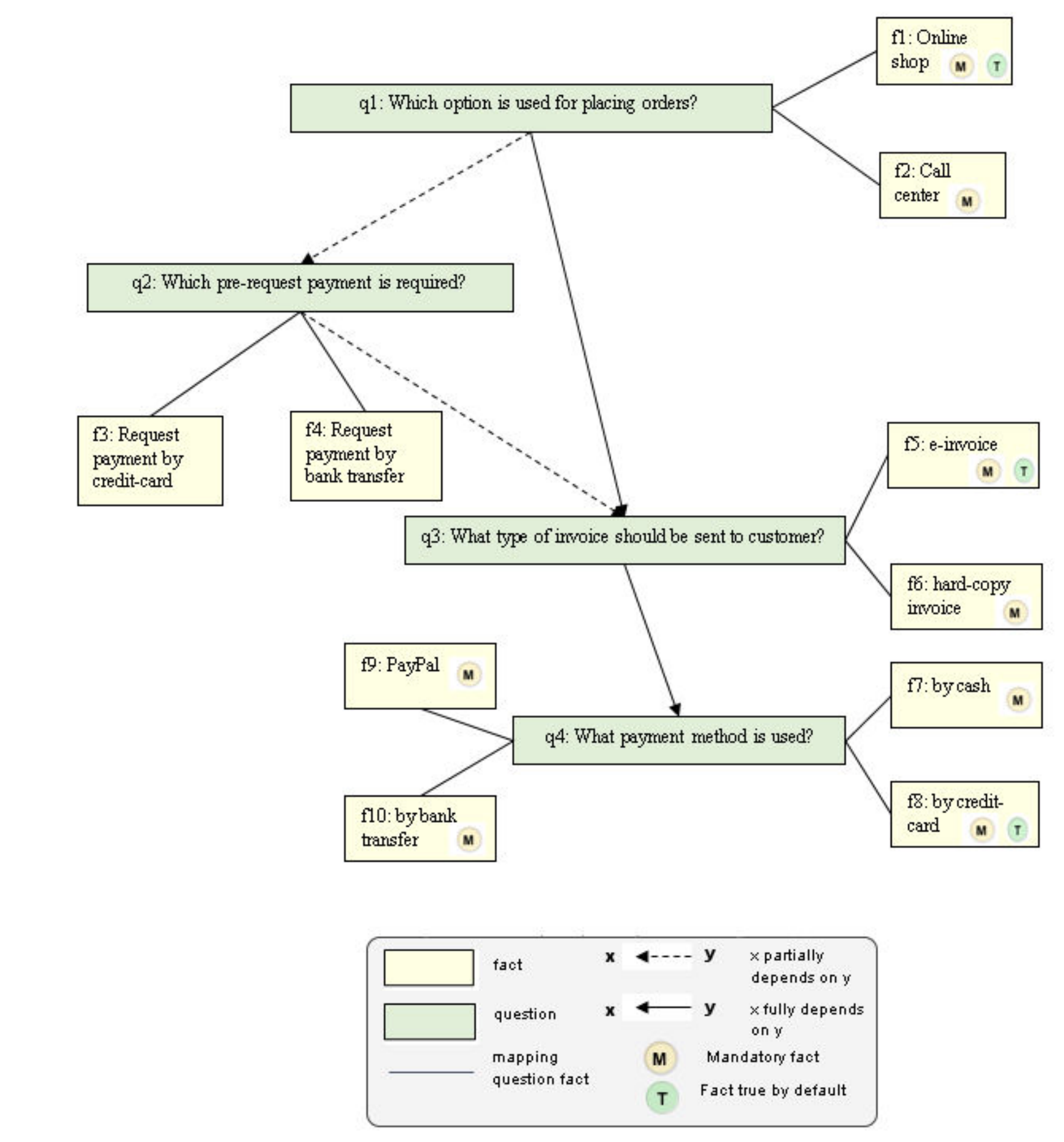}}		
	\caption[An extract of the questionnaire model]{An extract of the questionnaire model for configuring e-invoice type process model}
	\label{fig:NodeConfiguration3}	
\end{figure}

Instead, we design the questionnaire model based on their proposal to fit our example. It captures processing invoice payment properties as shown in Figure~\ref{fig:NodeConfiguration3} which comprise a set of features called domain facts organized into questions. All questions and facts have been assigned a unique identifier. A domain fact has a default value which is the most common choice for that fact, \eg , \emph{f5:e-invoice} as most of the invoice payments are e-invoice, then we can assign a false value to the other fact \emph{f6:hardcopy invoice}. Moreover, if a domain fact needs to be explicitly set when answering the question it is marked as mandatory. Otherwise, if a fact is left unset for the corresponding question then its default value can be used to answer the question or it is skipped. In a questionnaire model an order is established for posing questions to users in contrast with the feature model. This is achieved via order dependencies. There are two types of dependencies: \emph{full} and \emph{partial} dependency. For example, \emph{q2} can be posed after \emph{q1} is answered, this is expressed via the partial dependency between \emph{q1}, \emph{q2} depicted with a dashed line in Figure~\ref{fig:NodeConfiguration3}. Whereas, a full dependency, \eg , \emph{q4} is posed after \emph{q1}AND \emph{q3} is answered to capture the mandatory precedence in order to give priority to the most important questions.

After the clear overview of current process meta-modelling approaches we analyse and compare existing approaches by using different criteria.

\section{Comparative analysis of current approaches}
\label{MMfPV:compare}

Recently, some comparative studies have been reported in business process variability domain.
\citep{Rosa:2017:BPV} conducted a systematic inventory of approaches to customizable process modelling.
The authors identify and classify major approaches and provide a comparative evaluation with the objective to answer three research questions (such as represent common and distinct features of customizable process modelling approaches and research gaps exists in current Literature review (LR)).
Authors in \citep{AYORA2015} conducted a systematic literature review to evaluate existing variability support across all phases of the business process life cycle. The authors considered and categorized primary studies based on eight research questions (such as underlying business process modelling language used, tools available for enabling process variability, and validation of methods proposed). They developed a framework, called VIVACE to enable process engineers to evaluate existing process variability approaches. Then, they evaluate their framework against three main approaches from LR: C-EPCs, Provop, and PESOA. Our survey differs from theirs, as we restrict our search to only select those approaches (five out of twenty-eight papers from digital libraries) that introduce a meta-model for capturing variants of a business process. Some other work from authors in \citep{Valencca2013} focus on identifying not only characteristics of business process variability but also challenges in this field through a literature mapping study. But they didn't compare or analyse the surveyed approaches from LR.
Whereas, authors in \citep{Torres2012} give a comparison on their assessed approaches to make the produced process models artefacts more understandable to business analysts. Whereas, author in \citep{DOHRING2014} compare two approaches (C-YAWL and vBPMN)on the basis of a reference process model but using different types of
configuration and adaptation mechanisms. 
In contrast, we extended the approach from \citep{Rosa:2017:BPV} where we first describe each main
approach in detail (see \Cref{RelatedW_2}) secondly, applies an example to it, and last draws a comparative analysis based on evaluation of each criterion derived from our LR.\par

Table~\ref{tab:Comp1} summarizes the evaluation results for process variants meta-models approaches. Each column indicates to what extent the approach in question covers each evaluation criterion defined as follows. We used a "+" sign to indicate a criterion that is fulfilled, a "\textendash" sign to indicate a criterion that is not fulfilled, and a "+/\textendash" sign to indicate partial fulfilment. 
The first column lists the sixth main approaches including our approach. The next sixth columns indicate the coverage of each criterion. The last column indicate the modelling language(s) underlined by each approach.\\
\textbf{RQ1:} Which process types and process perspectives are covered by process variability meta-models?
From results of LR in respect to this research question we can conclude that 
two type of processes exist: design-time(\ie , variations is considered only during process modelling phase) and runtime (\ie , variations is considered only during process enactment for example to handle exceptions).
Whereas, process perspectives may categorized the surveyed approaches to mainly functional (what activities are captured) and behaviour perspectives (the control-flow sequence), even though some approaches deals somehow also with some aspects of organizational (resources to be consumed) and informational perspective (consumption of data).
Thereof, the criteria derived from \textbf{process types} results are: 
\vspace{-\topsep}
\begin{itemize}
\item \textbf{Conceptual}: If an approach is designed to support conceptual modelling only than this means that  variability is captured during process definition and these variant models will not be executed on top of a BPMS. Thereof, we say that this approach meets this criterion.
\item \textbf{Executable}: An approach meets this criterion if variability is considered for process models that are meant to be executed by a typical BPMS. Moreover, during their enactment there are no inconsistencies reported in associating between different elements of a process model.
\end{itemize}
\vspace{-\topsep}
Therefore, for \textbf{process modelling perspectives} results the criteria derived might be:
\vspace{-\topsep}
\begin{itemize} 
\item \textbf{Control-flow}: An approach meets this criterion if variability is captured along with activities and decision gateways that might become variation points (\eg , capture a skipped activity in one of the variants).
\item \textbf{Resources}: If the variability is captured in the participated resources (human or system) that are planned to perform different tasks. In so doing, resources can become variation points (\eg , a typical resource is not performing in some of the process variants). If the approach does not represent them graphically but it is only mentioned then we say that the approach partially fulfils this criterion \citep{Rosa:2017:BPV}.
\item \textbf{Data Objects}: An approach meets this criterion if data objects (\ie , produced-input data objects and consumed-output data objects) might become variation points.
For example, a pay invoice confirmation is not captured in one of the variants of a order-to-pay process.
If the approach does not represent them graphically but it is only mentioned then we say that the approach partially fulfils this criterion.\\
\end{itemize}
\vspace{-\topsep}
\textbf{RQ2:} Which supporting technique is used to introduce or capture variability between process models?
In respect to this research question, the criteria derived from \textbf{supporting technique} results are:
\vspace{-\topsep}
\begin{itemize}
\item \textbf{behavioural}: The approach takes as input a collection of process variants and derive a process variant by hiding and blocking process elements. Any behavioural anomalies such as deadlocks should be avoided.
\item \textbf{Structural}: The approach takes as input a base process model and after applying a set of change operations to it a process variant is derived. Any structural anomalies such as disconnected activities should be avoided.
\end{itemize}
\vspace{-\topsep}
According to the technique supported by specific approach some transformations should be done to process model in order to derive a variant. These transformations (by restriction/extension) might be categorized as criteria for:
\vspace{-\topsep}
\begin{itemize}
\item \textbf{Restriction}: An approach matches this criterion if a process model is configured
by restricting its behaviour.
\item \textbf{Extension}: An approach matches this criterion if a process model is configured
by extending its behaviour
\end{itemize}
\vspace{-\topsep}
\textbf{RQ3:} Which process is a specialization(generalization) of another process? The approach takes as input a collection of process variants and derive a process variant by applying substitution operations to its abstract activities.
Then, a valid criteria derived might be:
\vspace{-\topsep}
\begin{itemize}
\item \textbf{Process Specialization}: Specialization relationship between processes. An approach matches this criterion if a specialization/generalization relationship exists among processes.
\end{itemize}
\vspace{-\topsep}

\newcolumntype{L}[1]{>{\raggedleft\arraybackslash}p{#1}}
\newcolumntype{C}[1]{>{\centering\arraybackslash\hspace{0pt}}m{#1}}
\newcolumntype{V}[1]{>{\scriptsize\raggedright\hspace{0pt}}m{#1}}
\begin{table}[H]
\centering
\caption[Comparative analysis of approaches for business process variability]{Comparative analysis of approaches for business process variability management}
\label{tab:Comp1}
\arrayrulecolor[rgb]{0.502,0.502,0.502}
\begin{scriptsize}
\begin{tabular}{C{2cm}C{0.7cm}C{0.7cm}C{1cm}C{1cm}C{1cm}C{1cm}C{1cm}C{1cm}C{0.7cm}C{0.8cm}C{1.5cm}} 
\arrayrulecolor[rgb]{0.502,0.502,0.502}\toprule
 Main      Approaches &  \multicolumn{2}{m{1.8cm}}{Process Type}     &  \multicolumn{3}{m{3.3cm}}{Process Perspective}           &  \multicolumn{2}{m{1.8cm}}{Supporting Techniques}              & \multicolumn{2}{m{1.5cm}}{Variability Type}&  \multicolumn{1}{m{1.2cm}}{Process Specialization} & \multicolumn{1}{m{1.5cm}}{Process modelling Language } \\ 
\arrayrulecolor[rgb]{0.502,0.502,0.502}\toprule
                                                  & Conceptual                      & Executable   & Control-flow                            & Resources    & Objects      & Behavioural                                & Structural & Restriction                          & Extension & ~&  ~                              \\
\textbf{PESOA} \citep{Bayer05processfamily}                                             & +                                & \textendash    & +/\textendash                             & \textendash    & +            & \textendash                                 & \textendash  & +                                    & \textendash & \textendash & BPMN, UML ADs        \\
\textbf{BPFM} \citep{Moon2008}                                             & +                                & \textendash    & +/\textendash                             & \textendash    & \textendash    & \textendash                                 & \textendash  & +                                    & \textendash  & \textendash & UML ADs                 \\
\textbf{vBPMN} \citep{Dohring2011}                                             & +                                & +/\textendash  & +                                       & \textendash    & \textendash    & +                                         & +          & \textendash                            & +& \textendash  & Block-structured BPMN         \\
\textbf{PF} \citep{Kulkarni2011}                                                & +                                & \textendash    & +/\textendash                             & \textendash    & \textendash    & \textendash                                 & \textendash  & +                                    & \textendash & \textendash & BPMN                 \\
\textbf{C-iEPC} \citep{LAROSA2011}                                           & +                                & \textendash    & +                                       & +            & +            & +                                         & +          & +                                    & \textendash & \textendash & C-iEPC               \\
\textbf{PV Hierarchy} \citep{Berberi2021_1000141036}                                      & +                                & +    & +                                       & +    & \textendash    & +                                         & \textendash  & +                                    & \textendash & + & BPMN                 \\
\arrayrulecolor[rgb]{0.502,0.502,0.502}\toprule              
\end{tabular}
\arrayrulecolor[rgb]{0.502,0.502,0.502}
\end{scriptsize}
\end{table}
From the information of the above table we can conclude that all approaches covers the the conceptual level of process variants and the control-flow process perspective. Whereas, authors in \citep{LAROSA2011} captures variability in the participated resources (human or system)and data objects too. The supporting technique used to introduce or capture variability between process models is proposed by us, authors in \citep{LAROSA2011} and \citep{Dohring2011} and the latter proposed both \emph{restriction} and \emph{extension} to capture variability. And only the method in \citep{Berberi2021_1000141036} meets the criterion about \emph{process specialization}.
\section{Conclusions}
\label{PMMSumDisc1}
In this paper we summarized the list of approaches that have been proposed to explicitly and adequately capture and manage processes with variants.
As discussed, reference or customizable process models were introduced to model these variants collections in a way that each variant could be derived by
inserting/changing/removing an activity according to a process context. This survey reviewed current literature by providing an overview of meta-modelling approaches
that have been extended in order to capture the variations of business processes. Moreover, we ave a comparative analysis of these approaches based on different criteria we identified from this inventory.
A potential new area for future research could include in investigating the scalability of the proposed approaches to handle large and complex process variants.


\renewcommand{\bibname}{References}

\bibliography{reference_ER.bib} 
\bibliographystyle{plainnat}

\end{document}